\documentclass[structabstract]{aa}  
\usepackage{graphicx}
\usepackage{txfonts}
\usepackage{natbib}

\bibpunct{(}{)}{;}{a}{}{,} 

\begin{document}
   \title{The first INTEGRAL--OMC catalogue of optically variable sources 
\thanks{Available in electronic form at the CDS via anonymous ftp to cdsarc.u-strasbg.fr (130.79.128.5) or via \textit{http://cdsweb.u-strasbg.fr/cgi-bin/qcat?J/A+A/} and at at \textit{http://sdc.cab.inta-csic.es/omc/var/}}}


   \author{J. Alfonso-Garz\'{o}n
          \inst{1}
          \and
          A. Domingo\inst{1}
\and
J.M. Mas-Hesse\inst{1}
\and
A. Gim\'{e}nez\inst{2,1}
          }

   \institute{Centro de Astrobiolog\'{\i}a -- Departamento de Astrof\'{\i}sica
              (CSIC-INTA), POB 78, 28691 Villanueva de la Ca\~nada, Spain \\ \email{julia@cab.inta-csic.es}
         \and
              European Space Astronomy Centre (ESAC), European Space Agency, POB 78, 28691 Villanueva de la Ca\~nada, Spain\\
             }

   \date{Received 24 July 2012; accepted 28 September 2012 }


  \abstract
   {The Optical Monitoring Camera (OMC) onboard INTEGRAL provides photometry in
the Johnson V-band. With an aperture of 50 mm and a field of view of
5$^\circ$$\times$5$^\circ$, OMC is able to detect optical sources brighter than
$V\sim18$, from a previously selected list of potential targets of
interest. After more than nine years of observations, the OMC database
contains light curves for more than 70\,000 sources (with more than 50
photometric points each).}
   {The objectives of this work have been to characterize the potential variability of the
objects monitored by OMC, to identify periodic sources and to compute their periods,
taking advantage of the stability and long monitoring time of the OMC. }
   {To detect potential variability, we have performed a chi-squared
test, finding 5263 variable sources out of an initial sample of 6071 objects with good
photometric quality and more than 300 data points each. We have studied
the potential periodicity of these sources using a method based on the phase
dispersion minimization technique, optimized to handle light curves with very different shapes.}
   {In this first catalogue of variable sources observed by OMC, we provide for
each object the median of the visual magnitude, the magnitude at maximum and minimum brightness in the light curve during the window of observations, the period, when found, as well as the
complete intrinsic and period-folded light curves, together with some additional ancillary data.}
   {}


   \maketitle
%

\section{Introduction}

The study of variable stars is one of the principal areas of astronomical research. The
beginning of their study dates back to the 16th century when the first variable star, a
supernova, was identified, and shortly thereafter telescopes started to be used for
astronomy. Since then, an increasing number of variable stars were discovered and observed but it was with the upcoming of first photography and photoelectric photometry and later of the CCD
technology and multi-wavelength astronomy that the number of known variable stars
increased very rapidly. In 1948 the first edition of the General Catalogue of Variable
Stars (GCVS) was published containing 10\,820 stars. The GCVS has been updated regularly
until its last release \citep{Samus09}. The number of known variable stars has increased up to
around 200\,000 objects compiled in the Variable Star Index (VSX; \citealt{Watson06,Watson12}). The VSX is maintained by the American Association of Variable Stars Observers (AAVSO), and
contains all objects from the GCVS and the catalogues of the Northern Sky Variability
Survey (NSVS; \citealt{Wozniak04}), as well as data resulting from other automated
surveys, like the All Sky Automated Survey (ASAS; \citealt{Pojmanski97}), or the
catalogues compiled from the Optical Gravitational Lensing Experiment (OGLE; \citealt{Udalski92}) observations: the Miras and eclipsing binaries from OGLE-II data or the catalogue of variable stars from OGLE-III.

The advent of space astronomy has provided additional high quality data for
thousands of objects. The most significant impact has been
produced by the missions specifically dedicated to obtain photometric data of
high accuracy, in some cases during very long periods of time. While optimized
for astrometric studies, the Hipparcos/Tycho mission also produced high
accuracy photometric time series \citep{Perryman97, Hoeg97}. Even higher
photometric accuracy is provided by dedicated missions, aimed to study
asteroseismology or to search for exoplanetary transits, like the Canada's
Microvariability and Oscillations of STars (MOST; \citealt{Walker03}), the
Wide-field Infrared Explorer (WIRE; \citealt{Bruntt06}), the COnvection,
ROtation and planetary Transits (CoRoT) mission \citep{Auvergne09} or Kepler
\citep{Koch10}, which is yielding the highest accuracy light curves available up
to now.

Furthermore, spacecraft systems or secondary instruments on the satellites can provide very useful photometric information as well. \citet{Zwintz00} performed a detailed analysis of more than 4500 data sets from the Hubble Space Telescope (HST) guide star data, finding 20 stars to be variable. In a similar way, the observations performed by the spacecraft tracking cameras of Chandra have allowed to produce the Chandra Variable Guide Star Catalogue (VGUIDE) containing data of 827 stars for which accurate, long term photometry was available \citep{Nichols10}.

The International Gamma-Ray Astrophysics Laboratory (INTEGRAL) mission,
launched October 17th, 2002, \citep{Winkler03} includes a small optical
telescope, the Optical Monitoring Camera (OMC), to provide photometry in the
Johnson V-band simultaneously to the high-energy observations \citep{Mas03}. In
addition to monitoring the primary high-energy targets, OMC has been observing
serendipitously tens of thousands of potentially variable objects within its
field of view, preselected from existing catalogues of variable stars.

In this paper we present the first INTEGRAL--OMC catalogue of optically
variable sources (OMC--VAR hereafter) that contains photometric data in the V-band, as well as variability and periodicity data for more than 5000 objects observed from October 2002 to February 2010. The catalogue will be updated as new data are being
analysed. At the end of the mission we expect to produce the final catalogue
with variability information for more than 25\,000 objects monitored over a long
time period, with consistent and well calibrated photometry from space.

\section{The Optical Monitoring Camera onboard INTEGRAL}

The Optical Monitoring Camera (OMC) is a small refractive telescope with an
aperture of 50 mm focused onto a large format CCD (1024 $\times$ 2048 pixels)
working in frame transfer mode (1024 $\times$ 1024 pixels imaging area). With a
field of view of 5$^\circ$$\times$5$^\circ$ OMCprovides photometry in the
Johnson V-band (centred at 5500 \AA) and for sources from $V$ $\sim$ 7 mag (for
brighter sources saturation effects appear) to $V$ $\sim$ 16-17 mag (limit
magnitude for 3$\sigma$ source detection and 200~s integration time). Typical
observations are performed as a sequence of different integration times,
allowing for photometric uncertainties below 0.1 mag for objects with
V~$\le$~16. The accuracy is limited to 0.01 mag by the flat field
correction matrix. OMC was designed to observe the optical emission from the
prime targets of the gamma-ray instruments onboard INTEGRAL and has the same
field of view (FoV) as the fully coded FoV of the INTEGRAL X-ray monitor (JEM-X;
\citealt{Lund03}), and it is co-aligned with the central part of the gamma-ray instruments larger
fields of view, the IBIS imager \citep{Ubertini03} and the SPI spectrometer \citep{Vedrenne03}. The need to have a FoV as large as possible led to a rather large angular pixel size ($17\farcs5 \times
17\farcs5$), with a Gaussian Point Spread Function with FWHM $\sim$1.4 pix.

\subsection{The OMC Input Catalogue}

In addition to monitoring the INTEGRAL primary high-energy targets, OMC has the
capability to observe serendipitously around 100 sources within its FoV. These
targets have to be pre-selected on ground, since the limited telemetry available to OMC
(only $\sim 2.2$ kbps) does not allow to download full images to ground. These
additional targets are automatically selected from the OMC Input Catalogue (OMC--IC hereafter) by a
specific OMC Pointing Software (OMCPS) running at the INTEGRAL Science Operation Centre
(ISOC). Only $11\times11$ pix windows of the CCD containing those objects are transmitted
to ground. The OMC--IC was compiled before the start of the mission, and is described in
detail in \citet{Domingo03}. It has been updated several times to include new targets of
interest or to improve the accuracy in coordinates, especially of the high-energy
objects.

The current version of the OMC--IC contains over 500\,000 entries, namely:

\begin{itemize}
\item{astrometric and photometric reference stars,}
\item{known optical counterparts of gamma-ray sources,}
\item{known optical counterparts of X-ray sources,}
\item{point-like X-ray sources detected and catalogued by ROSAT,}
\item{quasars observable by the OMC,}
\item{known additional AGNs and }
\item{a big fraction of the known variable stars (including eruptive variable stars, novae and cataclysmic stars).}
\end{itemize}

The OMC--IC includes both known variable stars as well as suspected variable stars.
Special care was devoted to the compilation of extragalactic objects (AGN, QSO,
starburst galaxies, narrow emission line galaxies, etc.) as they are potential
high-energy emitters, and in many cases show significant variations of their optical flux. Indeed, \citet{Beckmann09} compiled the second INTEGRAL AGN catalogue including optical information from OMC for 57 AGN.

In the compilation of optical sources, mostly available before the INTEGRAL launch, the SIMBAD 
database and the following catalogues were used:


\begin{enumerate}

\item {\it Combined General Catalogue of Variable Stars, 4.1} (GCVS;
\citealt{Kholopov98}).\\ It contains a catalogue of all known Galactic variable stars
prior to 1997, a catalogue of extragalactic variable stars, a catalogue of supernovae and
a catalogue of suspected variable stars not designated as variables prior to 1980
(published in {\it The New Catalogue of Suspected Variable Stars}, \citealt{Kukarkin82}).

\item {\it New Catalogue of Suspected Variable Stars. Supplement}
      \citep{Kazarovets98}.\\
      It contains 11\,206 stars suspected of variability which were not
      designated as variables prior to 1997.

\item {\it The 74th Special Name-list of Variable Stars} \citep{Kazarovets99}.\\
      It contains 3157 variable stars whose variability was
      discovered by the Hipparcos mission. All the stars satisfy the
      GCVS variability criteria.

\item {\it The Hipparcos Catalogue} \citep{Perryman97}. \\
      The 11\,597 variable stars in the Hipparcos catalogue
      were included.

\item {\it Variable stars in the Tycho photometric observations}.\\
      To search for variability among faint stars of the Tycho catalogue \citep{Hoeg97}, 
      \cite{Piquard01} made a treatment that took 
      into account truncated detections and censored measurements. 
      The list contains 1091 stars suspected to be variable stars.

\item {\it Quasars and Active Galactic Nuclei, 8th Ed.} \citep{Veron98}.

\item {\it The Active Galactic Nuclei Catalogue} (Padovani priv. comm.). \\
      This catalogue includes 12\,021 quasars and active galaxies and is
      heavily based on the catalogue of {\it Quasars and Active Galactic
      Nuclei, 7th Ed.} \citep{Veron96}. It also includes the {\it BL Lac
      Catalogue} \citep{Padovani95} updated with BL Lac$'$s discovered
      in 1996, and the radio galaxies in the 1 Jy, S4, and S5 radio catalogues.
      The AGN catalogue reports $V$ magnitudes almost for all of their objects.
      Nevertheless, in some cases $V$ magnitudes were derived from $B$
      magnitudes by assuming a \hbox{$(B-V)$} colour index.

\item {\it Narrow Emission Line Galaxies} (Kunth priv. comm.).\\
      Compilation of 441 objects dominated by
      intense starburst activity. These objects show strong
      optical emission lines and/or very blue stellar continuum,
      both being tracers of young, massive stellar populations and therefore
      candidates for host supernovae.   

\item{\it Catalogue of Cataclysmic Variables} \citep{Downes01}.
      This catalogue contains 1134 sources that were included in the OMC-IC.

\begin{figure*}
\includegraphics[width=0.48\textwidth]{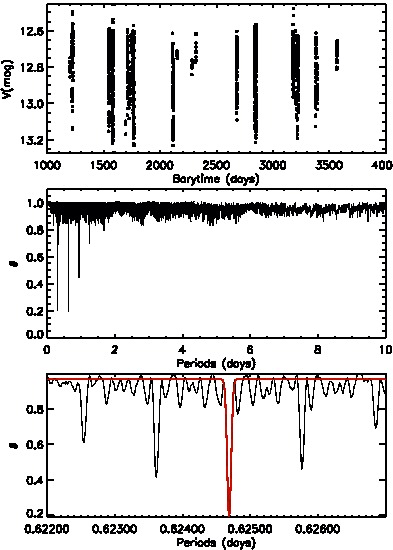}
\includegraphics[width=0.48\textwidth]{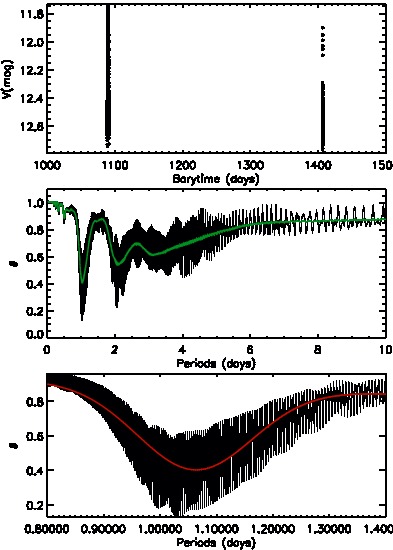}
\caption{Visual description of the process of period determination for two different
cases: a standard situation (analysis of IOMC 0460000022) on the left and an example of
aliasing (IOMC 5228000021) on the right. In both cases, the plots are: \textbf{Top: } The light curve without folding. For the light curve affected by aliasing the large temporal gap can be appreciated.
\textbf{Middle:} Periodogram from 0 to 10 days. The peaks are much narrower in the
standard case. In the aliasing example, the smoothed periodogram taking
into account the aliasing effect (see text) is overplotted in green. \textbf{Bottom:} Detail of the
method used to determine the uncertainty of the period. In red, the continuum plus gaussian fit to the peak is plotted.}
\label{fig1}
\end{figure*}

\item{\it Candidates for classical T-Tauri stars}\\
      Classical T Tauri stars are good targets for OMC because they emit in the optical
      band in a periodic or non periodic way, but always have some type of variability
      related to their physical conditions. The amplitudes of the light curves are in the
      range 0.01--3 mag. Periods are often ${\sim}1$ day, so the monitorization from a space platform should overcome the lack of observing them from the ground.
      732 candidates for classical T-Tauri stars were selected \citep{Caballero-Garcia03}
      based on the segregation shown by these stars in a 2MASS $(J-H)$ vs. $(H-Ks)$ 
      colour-colour diagram \citep{Lee02}.

\item{\it Candidates for cool dwarfs of G-K-M spectral type}.\\
      It is broadly known that cool dwarfs of G-K-M spectral type can be variable at a
      certain degree (mainly in the range 0.05--0.1 mag. and not greater than 0.2 mag.),
      showing very interesting properties of the behaviour of this kind of stars.
      A selection of 35\,101 candidates was made by \cite{Caballero-Garcia03} based on proper
      motions and the 2MASS $(J-H)$ and $(H-Ks)$ colours.

\end{enumerate}

\section{Data analysis}\label{analysis}


After the proprietary period of one year, all INTEGRAL data are open to the scientific
community.  At the moment of writing, INTEGRAL has been in orbit for more than nine
years and the OMC database (\textit{http://sdc.cab.inta-csic.es/omc/};
\citealt{Gutierrez04}) now contains observations for more than 130\,000 objects, of which more
than 70\,000 have light curves with at least 50 photometric data points. The data included in this catalogue contain observations from the beginning of the mission until February 2010. 

\subsection{Selection of the sources and cleaning of the light curves}

For this work, sources with more than 300 photometric points have been selected from the
OMC database, in order to deal with light curves with enough points to study their
variability and when possible, their periodicity. With the aim of improving the quality of
the data some selection criteria have been applied to the individual photometric points.

First of all, we have checked if the coordinates in the OMC--IC agree with the coordinates
given in SIMBAD. The OMC database allows the user to select between two different centroid
methods to extract the photometry, which corresponds to different options in the standard
OMC Off-line Science Analysis (OSA) software distributed by the INTEGRAL Science Data
Centre (ISDC; \citealt{courvoisier03}). If \emph{Source coordinates} (default) is
selected as the centroid method, the pipeline looks for the centroid allowing a maximum
offset of 10 arcsec with respect to the given coordinates. If the centroiding algorithm
fails by any reason, the extraction mask is nevertheless re-centered right on the source
coordinates as given in the OMC--IC. However, if \emph{brightest pixel} is
selected as the centroid method, the position of the extraction mask is re-centered on the
brightest pixel within the central $5\times5$ pixel section. For the sources with a difference
between the coordinates of the input catalogue and the coordinates given by SIMBAD greater
than 10 arcseconds, the brightest pixel method has been used. Moreover, in all the cases,
those photometric points with a distance to the coordinates of SIMBAD greater than 10
arcseconds have been rejected.

\begin{figure}
\resizebox{\hsize}{!}{\includegraphics{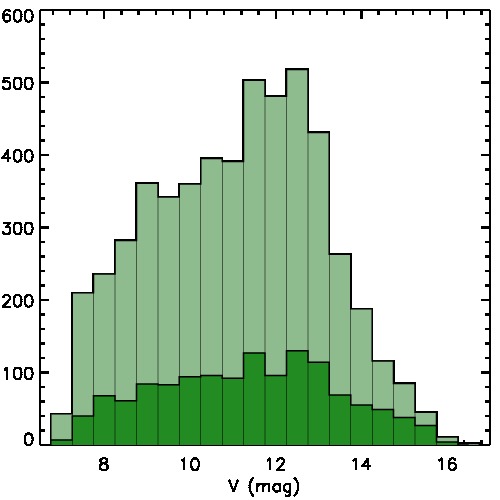}}
\caption{Histogram of magnitudes, covering the range from 7.10 to 16.27, with a peak of the
distribution at around 12.3 mag. In this and subsequent histograms the fraction of sources 
for which we have determined a period in this work are represented with dark green.}
\label{histo_magnitudes}
\end{figure}

Secondly, we have looked for duplicated entries in the catalogue. There are sources in the input
catalogue with different OMC identifiers, but with the same SIMBAD name or very close
coordinates. These two situations have been taken into account and the duplicated sources
have been removed from the list. 

Typical OMC integration times are 10~s, 50~s and 200~s, although at the begining of the
mission shots of  10~s, 30~s and 100~s were obtained. Therefore, photometric points with
different integration times can be found in the database. The longest integration times
led to saturation of the brightest objects. For this reason, for each light curve we have
rejected the photometric points that could be saturated. In general,  all the photometric
points within light curves where the median brightness is less than 7.1 were rejected. When
the median brightness was less than 8.3, those photometric points with exposure times
greater than 20~s were eliminated. If the median brightness was between 8.3 and 8.8, only
the photometric points with exposure times of 10 or 30~s were selected. For light curves
with median brightness between 8.8 and 9.6 the exposures of 100 and 200~s were rejected.
When the median brightness was between 9.6 and 10.3, the photometric points with exposure
times of 200~s were removed. And finally, for those sources with median brightness greater
than 10.3 we considered that no problems of saturation should appear and all photometric
points were considered. 

In a similar way, some detection limits were defined to avoid photometric points with too
low signal-to-noise ratio (SNR). For each light curve, those photometric points with exposure times of 10~s and a median brightness greater than 12, exposure times of 30~s and a median brightness
greater than 13 and exposure times of 50~s and a median brightness greater than 14 have been
removed.

Furthermore, measurements with low SNR were excluded. A minimum
SNR of 10 was required for the shortest integrations (10, 30 and 50~s) while for the
longest ones (100 s and 200 s) a criteria of SNR~$>3$ was imposed.

The observations were filtered from the effects of cosmic rays hits and some occasional
detector artifacts. Each photometric point was compared with the 20 closest ones
within 3 days (to avoid the timing gaps). For these 20 points, the standard deviation was computed and a 5-sigma rejection criteria was applied. We have checked manually
that this algorithm removed most of the outliers. Finally, the photometric points that
contained any hot pixel within their extracting mask of $3\times3$ pixels were
removed.

In this first version of the OMC catalogue we have analysed only those objects with a
given SIMBAD object type in the OMC--IC. Applying these filters we obtained 6071 sources with high-quality light curves to be studied. In the next version of the OMC catalogue we will
include all objects in the OMC database complying with the previous requirements,
independently of their assigned type, estimated to be around 25\,000.

\subsection{Detection of variability}
To detect which light curves showed variability we have used the $\chi^2$ goodness-of-fit test. This test compares a null hypothesis to an alternative hypothesis. If the value obtained for the test statistic is greater than a value corresponding to the chosen significance level of the $\chi^2$ distribution with k = n-1 degrees of freedom, being n the number of data points, then we reject the null hypothesis. We have used for each source the null hypothesis of being non-variable and having a constant magnitude equal to the mean. The alternative hypothesis is that the star is variable. Then, we calculated the $\chi^{2}$ and the significance $\alpha$. This significance gives the probability of being wrong when rejecting the null hypothesis (the source is constant). We have considered as variable those sources with $\alpha < 0.05$ (a lower limit probability of being variable of 95\%). Following this criterion we have identified 5263 sources showing variability in their OMC light curves, which constitute our present catalogue. Though all 6071 sources were expected to be variable, we have not identified variability for 808 of them, according to the criteria we have applied. In some cases this can be due to the time span of the OMC observations, and in other cases the amplitude is likely below the OMC variability detection
threshold used for this first version of the catalogue.
%

\subsection{Study of the periodicity}\label{studyP}

To determine which sources are periodic and to derive their periods, an algorithm based on
the Phase Dispersion Minimization (PDM) technique \citep{Stellingwerf78} has been
developed. Given the variety of variability patterns present in our data we decided not to
use the Lomb-Scargle (L-S) periodogram technique by \cite{Scargle82}. This method evaluates the
discrete Fourier transform for nonevenly sampled data. Since this is equivalent to assuming a
prior sinusoidal light curve, we considered it not adequate for most eclipsing and
non-sinusoidal light curves.

\begin{figure}
\resizebox{\hsize}{!}{\includegraphics{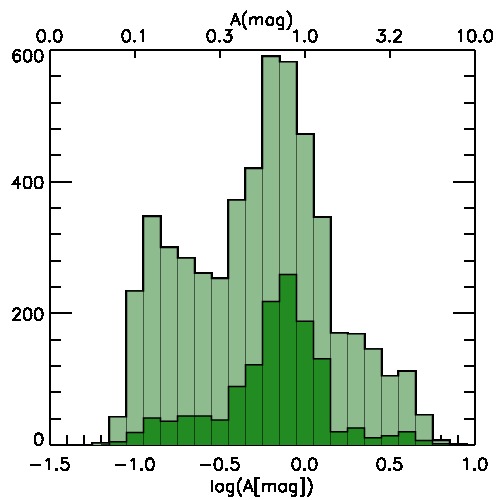}}
\caption{Histogram of amplitudes (calculated as the difference between the magnitude at minimum and the magnitude at maximum of the light curve) with a peak of the distribution at around 0.84 mag. In dark green are represented the sources for which we have determined a period in this work.}
\label{histo_amplitudes}
\end{figure}

We explored a wide range of frequencies corresponding to periods from 0.8 h (a lower
limit of the periods that can be detected by OMC, considering the typical exposure times),
up to half of the time range of the observations (the maximum period to have at least
two cycles in the folded light curve). We first used a frequency increment of
$\frac{1}{3B}$, being $B$ the total time baseline of the observations, and in a second
iteration around the first trial period, we used a period increment of $2 P^{2} /(n B)$
\citep[see][]{Gimenez83}, where $P$ is the trial period and $n$ is the number of points in the
light curve. For each period, the time-folded light curve is divided into a series of bins
and the variance of the amplitude within each bin with respect to a mean curve is
computed. In our algorithm the number of bins is variable and it depends on the quality of
each light curve. For each bin, the mean value is calculated and linear interpolations are
done between the means in order to obtain one average curve for each folded light curve.
The ratio between the sum of the bin variances and the overall variance of the data set is
called $\Theta$, and the period that minimizes this parameter has been considered to be
the best estimate.

\begin{figure}[ht]
\resizebox{\hsize}{!}{\includegraphics{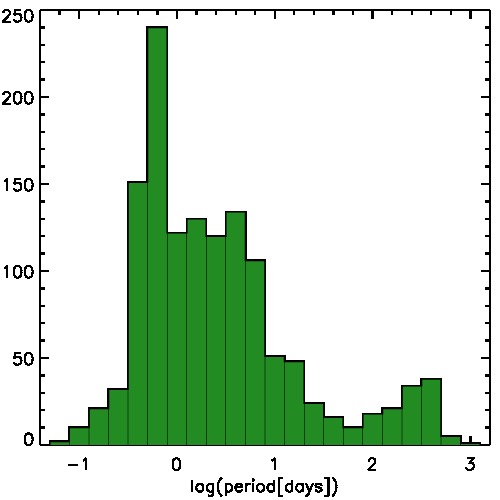}}
\caption{Histogram of the periods derived. Typical values vary between a few hours and 10
days, with a peak of frequency at around 15 h.}
\label{histo_periods}
\end{figure}

\begin{table*}
\caption{Contents of the OMC--VAR catalogue, accessible at \textit{http://sdc.cab.inta-csic.es/omc/var/} and at CDS services.}
\label{table1} 
\centering 
\begin{tabular}{l c c} 
\cline{1-3} 
\cline{1-3} 
Field & Source & Description \\ 
\cline{1-3} 
IOMC & OMC & OMC Identifier \\ 
SIMBAD name & SIMBAD & SIMBAD Basic Identifier \\ 
RA & SIMBAD &  Right Ascension in sexagesimal units (J2000) \\ 
DE & SIMBAD &  Declination  in sexagesimal units (J2000) \\ 
RAdeg & SIMBAD & Right Ascension in degrees (J2000) \\ 
DEdeg & SIMBAD & Declination (J2000)  in degrees (J2000)  \\ 
otype & SIMBAD & SIMBAD object type \\ 
sptype & SIMBAD &  SIMBAD spectral type \\ 
vartype & VSX & VSX variability type \\ 
vargroup & VSX, SIMBAD, others & General Variability Group \\ 
V mag & OMC & Median V magnitude \\ 
magerr & OMC &  Mean of the magnitude error \\ 
magMax & OMC &  Magnitude at maximum brightness \\ 
magMin & OMC &  Magnitude at minimum brightness\\ 
per & OMC & Period of variability in days \\ 
errper & OMC & Uncertainty of the period in days  \\ 
timebase & OMC & Time base of the observations in days  \\ 
tagcont & UCAC3, NOMAD1, SIMBAD, VSX & Tag of contamination \\ 

\cline{1-3}
\end{tabular}
\end{table*}

\begin{figure}
\resizebox{0.85\hsize}{!}{\includegraphics{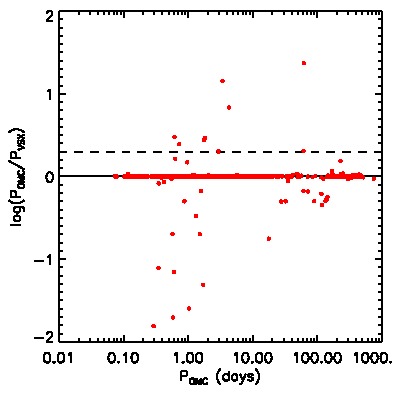}}
\caption{Comparison of periods derived from OMC data (this work) with the values compiled by the VSX. The solid line represents $P_{OMC} = P_{VSX}$  while the dashed line marks the condition $P_{OMC} = 2 P_{VSX}$.}
\label{comp-periods}
\end{figure}

With the values of the trial periods and their corresponding values of $\Theta$, we
computed the periodograms. The uncertainty in the period was calculated by fitting a gaussian to the peak corresponding to the minimum value of $\Theta$ in the periodogram. The 1-sigma error in the determination of the centroid of the gaussian is provided for each source as the standard uncertainty of the period and corresponds to a confidence interval of 68.3\% (see Fig.~\ref{fig1}). For most of the sources the process above works fine, but in some cases aliasing becomes a problem. Aliasing occurs when a curve is observed with an insufficient sampling. For some sources in the catalogue there are temporal gaps in the observations much greater than the value of the period, and the time intervals of the different observations are not long enough to determine the period accurately. For a given estimated period $P_{i}$, if the temporal gap, $t_\mathrm{gap}$ (the time when the source has not been observed), is such as
$t_\mathrm{gap} = N \times P_{i}$ (with $N$ big enough), there will be many significant
peaks in the periodogram, close to the deepest one, corresponding to that values of the
period that make $t_\mathrm{gap} = (N+1) \times P_{i-1}$ or $t_\mathrm{gap} = (N-1) \times
P_{i+1}$ (see Fig.~\ref{fig1}, right panel). In these cases we have smoothed the
periodogram with a window size wider than the distance between these false peaks. This smoothed
curve is used to determine the limit of significance of the peak and the half width of this
peak is used as initial value of sigma in the gaussian fit. Finally, a visual inspection
of each folded light curve was performed to identify and correct potential problems due to
ambiguities in the results, such as deriving the half or double period in some eclipsing
binaries or pulsating stars with symmetrical light curves. 

The method used in this work yields only the primary periods. To obtain second or multiple periods,  if present, a manual analysis of each source is required and this is out of 
the scope of this paper.

We show in Fig.~\ref{period1} the light curve of Her X--1, one of the brightest and most
studied persistent X-ray binary pulsars. The system displays a great variety of
phenomena at different timescales, including pulsations at 1.24 s, eclipses at the orbital
period of 1.7 days, and a 35 day cycle in the X-ray intensity \citep{Leahy10}. Despite the
large gap  between the two epochs at which this object has been observed by OMC, our
method retrieves properly the orbital period. Nevertheless, the length of the gap, compared
to the short orbital period, implies that the period has been derived with a rather low
accuracy.

\subsection{Photometric contamination by nearby stars}\label{cont}

For the data analysed in this study, fluxes and magnitudes were derived from a photometric
aperture of 3$\times$3 pixels (1 pixel = $17\farcs504$), slightly circularized, i.e.,
removing $\frac{1}{4}$ of a pixel from each corner (standard output from OSA). Therefore
the computed values include the contributions by any other source inside the photometric
aperture.  We flagged in a column of the catalogue those sources that might be affected by
this kind of photometric contamination. Due to the variable nature of the sources in the
catalogue and the difficulty to accurately estimate the contamination by the nearby and
background stars, we developed three different methods. The first one was to take into
account the photometric information of the sources close to the extraction aperture
present in the UCAC3 \citep{zacharias10} and NOMAD1 \citep{Zacharias04} catalogues. The
second one was to compare the $V$ magnitude provided by SIMBAD (for variable sources it is
usually the magnitude at maximum brightness) with the lowest magnitude we measure for each source,
and the third one was to compare the median magnitude in VSX with the median magnitude we
measure. We considered a source significantly contaminated when the measured flux was a
factor of 2 higher than expected.  This is equivalent to a difference between expected and
observed magnitude of more than 0.75 mag. While this criterion is not very restrictive in
terms of photometric accuracy, it allows to study variability and search for periodicity
for objects with a significant degree of contamination. We want to stress nevertheless
that this procedure has the caveat that in some cases the variability/periodicity found
might be originated by the contaminating object(s), and not by the target source. One of the objects mostly affected by blending is HH Nor, shown in Fig.~\ref{period2}, whose variability
pattern is completely dominated by the variability of the brighter, nearby object VSX
J154329.4--515037, which is indeed a variable RR Lyr star with a much shorter period.
Being located at just $12\arcsec$, a fraction of an OMC pixel, the intrinsic variability
of this contaminating star completely hides the variability of the target source, HH Nor.

The results of these three methods are coded in the column of contamination with three
numbers  (a reference to each method) where 1 means that the source is not contaminated
(under the conditions defined above), 2 means that it is contaminated and 3 means no
photometric information has been found for that source. Because of the inherent
difficulties of estimating the contamination of variable sources, this flag is for
guidance only and to study individual sources a more exhaustive analysis of the sky field
around each source is recommended.

\section{Contents of the OMC--VAR catalogue}

\begin{figure*}
\centering
\includegraphics[width=0.8\textwidth]{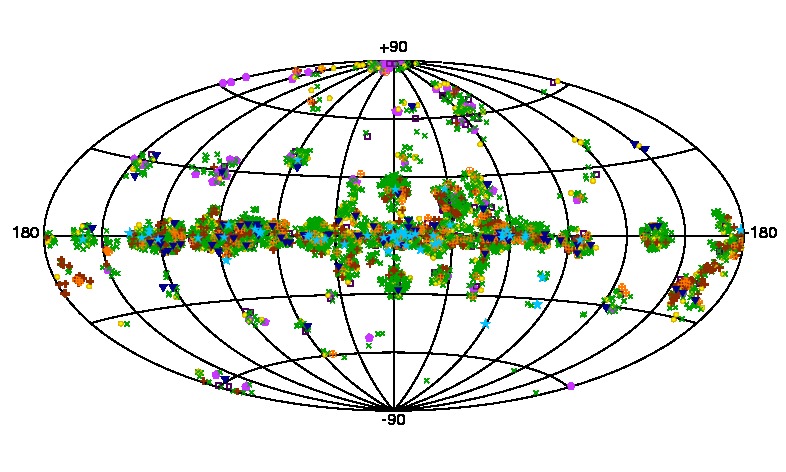}
\caption[]{\footnotesize{Distribution in galactic coordinates of all the sources in the
catalogue. The green crosses represent pulsating stars, the red filled points correspond
to eclipsing binaries, the brown plusses are eruptive stars, the orange complex plusses
represent rotating stars, the inverted dark blue filled triangles represent cataclysmic
variables, the light blue filled stars are X-ray binaries, the yellow filled points
correspond to objects simply classified as variable stars, the purple filled pentagons represent
extragalactic objects and the empty purple squares are other types of objects.}}\label{mapa}
\end{figure*}

Table \ref{table1} describes the contents of the OMC--VAR catalogue, which is fully accessible 
online. In this first release we provide information about the variability of 
5263 objects distributed all over the sky (see Fig.~\ref{mapa}). The 
majority of the objects are located in the Galactic plane and Galactic 
bulge, as well as on specific areas of the sky. This distribution reflects 
the integration time density maps of INTEGRAL, driven by the location of the 
most interesting high-energy sources.

For each source we include the median value of the $V$ magnitude, the mean of the
photometric errors of the points and the magnitudes at maximum and minimum brightness, estimated as the 2nd and 98th percentiles, respectively, of the points in the light curve. For those sources
classified as periodic or probably periodic, we provide the best estimate of the period
and its calculated uncertainty.

\begin{figure}
\resizebox{0.9\hsize}{!}{\includegraphics{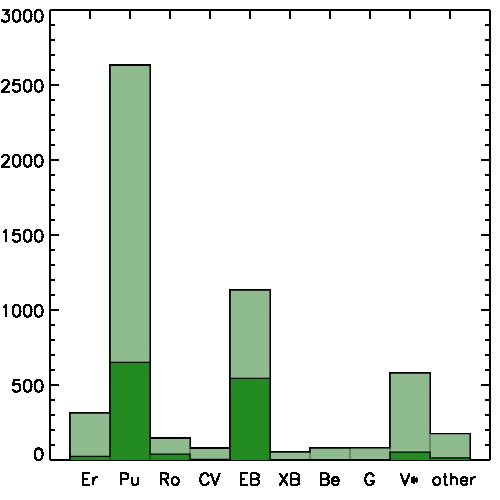}}
\caption{Histogram of the different groups of variability types present in the catalogue. There are
a big number of objects which are unclassified, or just classified as variable objects. }
\label{hist_otypes}
\end{figure}

The values of the median magnitudes go from 7.10 mag to 16.27 mag. In Fig.~\ref{histo_magnitudes}
we show the histogram of the median magnitudes, showing a peak at around 12.3 mag. A similar plot for the amplitudes (calculated as the difference between the magnitude at minimum and the magnitude at maximum of the light curve) is shown in Fig.~\ref{histo_amplitudes}. The amplitudes go from 0.06 mag to 7.41 mag with a mean value of 0.84 mag. 

The time base for each light curve is also provided. These time bases have a minimum value of 0.69 d, a maximum value of 2815.19 d and a mean value of 1862.0 d.

We have determined  periods for 1337 sources, out of the 5263 variable objects compiled in
the catalogue. The distribution of periods is shown  in Fig.~\ref{histo_periods}. Typical
values vary between a few hours and 10 days, with a peak of frequency at around 15 h.
Some sources present periods of up to some hundreds of days, but there are not many of
them given the time span and distribution of the observations. The VSX database also
provides information about the periodicity of the sources and we have used this
information to compare with our results. We have derived periods for 175 objects
whose periodicity was unknown and in many other cases, we have improved the results with
respect to those found in the VSX database, as can be seen in Appendix~A.

In Fig.~\ref{comp-periods} we compare the periods derived from OMC data in this work with
the values compiled in the VSX, when both were available. For the large majority of
objects the values are consistent, though we want to stress that in many cases, even a
very small improvement on the determination of the period is enough to yield a
satisfactory folding of the light curve (see the examples in Appendix~A). A dashed line
marks the points for which  $P_{\rm OMC} = 2 \times P_{\rm VSX}$, to identify those cases of binaries with eclipses of similar shape that could have been misidentified in the VSX. In the rest
of the cases we have verified that the period given in the VSX was not consistent with the
OMC light curves. This happened sometimes for systems with multiple periods due to different variability mechanisms like rotation, orbital, X-rays, etc. One example is Her~X--1 for which, as we mentioned in section \ref{studyP}, the period given by VSX does not correspond to the orbital one to which OMC is more sensitive (see Fig.~\ref{period1}).   

\subsection{Classification}

\begin{figure}
\resizebox{0.8\hsize}{!}{\includegraphics{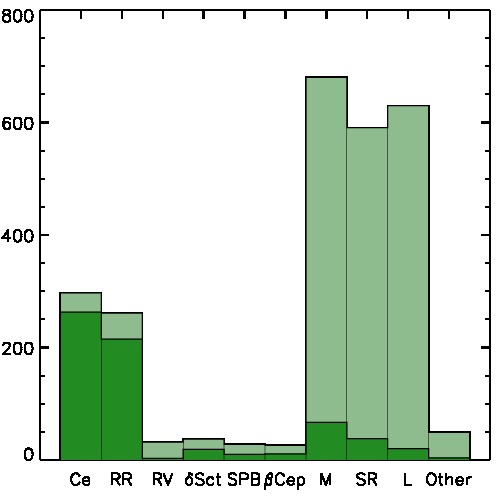}}
\resizebox{0.8\hsize}{!}{\includegraphics{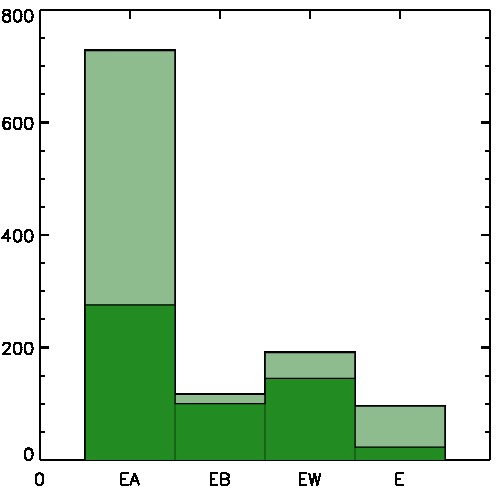}}
\caption{Histogram of the subgroups of variability types present in the catalogue.
\textbf{Top:} Histogram of the different kinds of pulsating stars. \textbf{Bottom:} 
Histogram of the subtypes of eclipsing binary stars.}
\label{hist_object_subtypes}
\end{figure}

A big variety of variable objects have been observed by OMC and are present in this
version of the catalogue. In Fig.~\ref{mapa} the distribution in galactic coordinates of the different kind of objects is plotted. In order to do a statistical study of the type of objects in the catalogue we have used the information available in the literature. When available, the
object classification has been extracted from the VSX.
Otherwise, the type of variability has been taken from the SIMBAD database. According to
the above classifications we have defined different variability groups and have assigned
each source to one of them. In the case of pulsating stars we have defined 10 additional
subgroups, and in the case of eclipsing binaries we have included 4 additional subgroups,
based only on the shape of the light curve: 

\begin{itemize}
\item 2631 pulsating stars (Pu), including 

\subitem {\textbullet} 297 Cepheids (Ce), 
\subitem {\textbullet} 261 RR Lyrae (RR) pulsators, 
\subitem {\textbullet} 37 $\delta$ Scuti pulsators ($\delta$ Sct), 
\subitem {\textbullet} 32 RV Tau stars (RV), 
\subitem {\textbullet} 28 $\beta$ Cephei stars ($\beta$ Cep), 
\subitem {\textbullet} 26 SPB pulsators (SPB), 
\subitem {\textbullet} 680 Mira type stars (M), 
\subitem {\textbullet} 591 semi-regular stars (SR), 
\subitem {\textbullet} 630 slow pulsators (L),
\subitem {\textbullet} 49 simply classified as pulsators (other).

\item 1132 eclipsing binaries (EB), including

\subitem {\textbullet} 728 Algol ($\beta$ Persei)-type eclipsing systems (EA), 
\subitem {\textbullet} 191  $\beta$ Lyrae-type eclipsing systems (EB), 
\subitem {\textbullet} 117 W Ursae Majoris-type eclipsing variables (EW),
\subitem {\textbullet} 96 simply classified as eclipsing (E)

\item 53 X-ray binaries (XB),
\item 78 cataclysmic variables (CV), 
\item  312 eruptive stars (Er), 
\item 144 rotating stars (Ro), 
\item 79 Be stars (Be), 
\item 579 stars just classified as variable (V*),
\item 80 extragalactic sources (G), and 
\item 162 objects of other types.
\end{itemize}

We show in Fig.~\ref{hist_otypes} the histogram corresponding to the different groups,
clearly dominated by the groups of pulsating stars and eclipsing binaries. In Fig.~\ref{hist_object_subtypes} we detail the histograms for the subgroups of pulsating stars and eclipsing binary systems, the latter dominated by Algol-type systems.  

Some examples of the kind of objects that can be found in the catalogue are shown in
Appendix A, where we have plotted some representative charts including the DSS image
around the target, and their light curves, unfolded and folded with the periods we have
derived and/or with the catalogued ones. 

\section{Summary}

We have analysed the optical photometric light curves, covering the time from the beginning of the
mission to February 2010, of the 6071 objects present in the OMC database  that
satisfy the criteria explained in section \ref{analysis}, with the aim of characterizing 
their expected variability and potential periodicity. 

In this first version of the INTEGRAL--OMC catalogue of optically variable sources we
have detected variability for 5263 objects, with median brightnesss in the range from
7.1 to 16.27 in $V$.

For 1337 of them we have been able to derive a period with enough significance. The periods found
vary from a few hours to some hundreds of days, with a peak in the histogram of typical
periods around 15 h. We inspected visually the light curves to check the validity of
the estimated periods, rejecting those false positives yielded by the PDM algorithm.

For each source in the catalogue we include the median value of the $V$ magnitude, the
mean of the photometric errors of the points and the minimum and maximum value of $V$,
estimated as the 98th and 2nd percentiles, respectively, of the points in the light curve. For
those sources classified as periodic or probably periodic, we provide the best estimate of
the period and its calculated uncertainty. We have also included a set of tags to identify
those sources whose photometry can be significantly affected by nearby stars, as well as
some ancillary data. 

This catalogue contains data useful for the study of several classes of variable stars:
eclipsing binaries, pulsating stars, active stars, cataclysmic variables, X-ray binaries,
etc. The most frequent objects in the present compilation are pulsating stars and
eclipsing binary systems, according to the classification provided by the VSX catalogue
and SIMBAD. A deeper study of some of the sources presented here is being carried out and
will be presented in forthcoming papers.

The next version of the INTEGRAL--OMC catalogue of optically variable sources will
include all other objects in the OMC database complying with the same photometric requirements,
estimated to be around 25\,000.

\begin{acknowledgements}

The development, operation and exploitation of OMC have been funded by Spanish MICINN
under grants AYA2008-03467 and AYA2011-24780 and previous ones.  
INTEGRAL is an ESA project funded by ESA member states (especially the PI countries:
Denmark, France, Germany, Italy, Spain, Switzerland), Czech Republic, Poland, and with the
participation of Russia and the USA. OMC was also funded by Ireland, United Kingdom, Belgium and the Czech Republic.
This research has made use of data from the OMC Archive at CAB (INTA-CSIC), pre-processed
by ISDC, and of the SIMBAD database, operated at CDS, Strasbourg, France.

\end{acknowledgements}

\bibliographystyle{aa} 
\bibliography{allcites}

\begin{thebibliography}{40}
\expandafter\ifx\csname natexlab\endcsname\relax\def\natexlab#1{#1}\fi

\bibitem[{Auvergne {et~al.}(2009)Auvergne, Bodin, Boisnard, Buey, Chaintreuil,
  Epstein, Jouret, Lam-Trong, Levacher, Magnan, Perez, Plasson, Plesseria,
  Peter, Steller, Tiph{\`{e}}ne, Baglin, Agogu{\'{e}}, Appourchaux, Barbet,
  Beaufort, Bellenger, Berlin, Bernardi, Blouin, Boumier, Bonneau, Briet,
  Butler, Cautain, Chiavassa, Costes, Cuvilho, Cunha-Parro, de~Oliveira~Fialho,
  Decaudin, Defise, Djalal, Docclo, Drummond, Dupuis, Exil, Faur{\'{e}},
  Gaboriaud, Gamet, Gavalda, Grolleau, Gueguen, Guivarc{'}h, Guterman, Hasiba,
  Huntzinger, Hustaix, Imbert, Jeanville, Johlander, Jorda, Journoud, Karioty,
  Kerjean, Lafond, Lapeyrere, Landiech, Larqu{\'{e}}, Laudet, Le~Merrer,
  Leporati, Leruyet, Levieuge, Llebaria, Martin, Mazy, Mesnager, Michel,
  Moalic, Monjoin, Naudet, Neukirchner, Nguyen-Kim, Ollivier, Orcesi, Ottacher,
  Oulali, Parisot, Perruchot, Piacentino, Pinheiro~da Silva, Platzer, Pontet,
  Pradines, Quentin, Rohbeck, Rolland, Rollenhagen, Romagnan, Russ, Samadi,
  Schmidt, Schwartz, Sebbag, Smit, Sunter, Tello, Toulouse, Ulmer, Vandermarcq,
  Vergnault, Wallner, Waultier, \& Zanatta}]{Auvergne09}
Auvergne, M., Bodin, P., Boisnard, L., {et~al.} 2009, \aap, 506, 411

\bibitem[{Beckmann {et~al.}(2009)Beckmann, Soldi, Ricci, Alfonso-Garz{\'{o}}n,
  Courvoisier, Domingo, Gehrels, Lubi{\'{}}nski, Mas-Hesse, \&
  Zdziarski}]{Beckmann09}
Beckmann, V., Soldi, S., Ricci, C., {et~al.} 2009, \aap, 505, 417

\bibitem[{Bruntt \& Buzasi(2006)}]{Bruntt06}
Bruntt, H. \& Buzasi, D.~L. 2006, \memsai, 77, 278

\bibitem[{Caballero-Garc{\'{i}}a(2003)}]{Caballero-Garcia03}
Caballero-Garc{\'{i}}a, M.~D. 2003, Master's thesis, Universitat de Barcelona

\bibitem[{Courvoisier {et~al.}(2003)Courvoisier, Walter, Beckmann, Dean,
  Dubath, Hudec, Kretschmar, Mereghetti, Montmerle, Mowlavi, Paltani,
  Preite~Martinez, Produit, Staubert, Strong, Swings, Westergaard, White,
  Winkler, \& Zdziarski}]{courvoisier03}
Courvoisier, T. J.~L., Walter, R., Beckmann, V., {et~al.} 2003, \aap, 411, L53

\bibitem[{Domingo {et~al.}(2003)Domingo, Caballero, Figueras, Jordi, Torra,
  Mas-Hesse, Gim{\'{e}}nez, Hudcova, \& Hudec}]{Domingo03}
Domingo, A., Caballero, M.~D., Figueras, F., {et~al.} 2003, \aap, 411, L281

\bibitem[{Downes {et~al.}(2001)Downes, Webbink, Shara, Ritter, Kolb, \&
  Duerbeck}]{Downes01}
Downes, R.~A., Webbink, R.~F., Shara, M.~M., {et~al.} 2001, \pasp, 113, 764

\bibitem[{Gim{\'{e}}nez \& Garc{\'{i}}a(1983)}]{Gimenez83}
Gim{\'{e}}nez, A. \& Garc{\'{i}}a, J.~M. 1983, Revista de la Real Academia de
  Ciencias Exactas, 1, 297

\bibitem[{Guti{\'{e}}rrez {et~al.}(2004)Guti{\'{e}}rrez, Solano, Domingo, \&
  Garc{\'{i}}a}]{Gutierrez04}
Guti{\'{e}}rrez, R., Solano, E., Domingo, A., \& Garc{\'{i}}a, J. 2004, in
  Astronomical Society of the Pacific Conference Series, Vol. 314, Astronomical
  Data Analysis Software and Systems (ADASS) XIII, ed. .~D.~E. F.~Ochsenbein,
  M. G.~Allen, 153

\bibitem[{Hoeg {et~al.}(1997)Hoeg, B{\"{a}}ssgen, Bastian, Egret, Fabricius,
  Gro{\ss{}}mann, Halbwachs, Makarov, Perryman, Schwekendiek, Wagner, \&
  Wicenec}]{Hoeg97}
Hoeg, E., B{\"{a}}ssgen, G., Bastian, U., {et~al.} 1997, \aap, 323, L57

\bibitem[{Kazarovets {et~al.}(1999)Kazarovets, Samus, Durlevich, Frolov,
  Antipin, Kireeva, \& Pastukhova}]{Kazarovets99}
Kazarovets, A.~V., Samus, N.~\., .~N., Durlevich, O.~V., {et~al.} 1999,
  Informational Bulletin on Variable Stars, 4659, 1

\bibitem[{Kazarovets {et~al.}(1998)Kazarovets, Samus, \&
  Durlevich}]{Kazarovets98}
Kazarovets, V., Samus, N.~\., .~N., \& Durlevich, O.~V. 1998, Informational
  Bulletin on Variable Stars, 4655, 1

\bibitem[{Kholopov {et~al.}(1998)Kholopov, Samus, Frolov, Goranskij, Gorynya,
  Karitskaya, Kazarovets, Kireeva, Kukarkina, Kurochkin, Medvedeva, Pastukhova,
  Perova, Rastorguev, \& Shugarov}]{Kholopov98}
Kholopov, P.~N., Samus, N.~N., Frolov, M.~S., {et~al.} 1998, in Combined
  General Catalogue of Variable Stars, 4.1 Ed (II/214A). (1998), 0

\bibitem[{Koch {et~al.}(2010)Koch, Borucki, Basri, Batalha, Brown, Caldwell,
  Christensen-Dalsgaard, Cochran, DeVore, Dunham, Gautier, Geary, Gilliland,
  Gould, Jenkins, Kondo, Latham, Lissauer, Marcy, Monet, Sasselov, Boss,
  Brownlee, Caldwell, Dupree, Howell, Kjeldsen, Meibom, Morrison, Owen,
  Reitsema, Tarter, Bryson, Dotson, Gazis, Haas, Kolodziejczak, Rowe,
  Van~Cleve, Allen, Chandrasekaran, Clarke, Li, Quintana, Tenenbaum, Twicken,
  \& Wu}]{Koch10}
Koch, D.~G., Borucki, W.~J., Basri, G., {et~al.} 2010, \apjl, 713, L79

\bibitem[{Kukarkin \& Kholopov(1982)}]{Kukarkin82}
Kukarkin, B.~V. \& Kholopov, P.~\., .~N. 1982, New catalogue of suspected
  variable stars (Moscow: Publication Office "Nauka")

\bibitem[{Leahy \& Dupuis(2010)}]{Leahy10}
Leahy, D.~A. \& Dupuis, J. 2010, \apj, 715, 897

\bibitem[{Lee \& Chen(2002)}]{Lee02}
Lee, H.~T. \& Chen, W.~P. 2002, in 8th Asian-Pacific Regional Meeting, Volume
  II, ed. S.~Ikeuchi, J.~Hearnshaw, \& T.~Hanawa, 161--162

\bibitem[{Lund {et~al.}(2003)Lund, Budtz-J{\o}rgensen, Westergaard, Brandt,
  Rasmussen, Hornstrup, Oxborrow, Chenevez, Jensen, Laursen, Andersen,
  Mogensen, Rasmussen, Om{\o}, Pedersen, Polny, Andersson, Andersson, K\"
  am\"~ar\" ainen, Vilhu, Huovelin, Maisala, Morawski, Juchnikowski, Costa,
  Feroci, Rubini, Rapisarda, Morelli, Carassiti, Frontera, Pelliciari,
  Loffredo, Mart{\'{i}}nez N{\'{}}~u\ nez, Reglero, Velasco, Larsson, Svensson,
  Zdziarski, Castro-Tirado, Attina, Goria, Giulianelli, Cordero, Rezazad,
  Schmidt, Carli, Gomez, Jensen, Sarri, Tiemon, Orr, Much, Kretschmar, \&
  Schnopper}]{Lund03}
Lund, N., Budtz-J{\o}rgensen, C., Westergaard, N.~J., {et~al.} 2003, \aap, 411,
  L231

\bibitem[{Mas-Hesse {et~al.}(2003)Mas-Hesse, Gim{\'{e}}nez, Culhane, Jamar,
  McBreen, Torra, Hudec, Fabregat, Meurs, Swings, Alcacera, Balado, Beiztegui,
  Belenguer, Bradley, Caballero, Cabo, Defise, D{\'{i}}az, Domingo, Figueras,
  Figueroa, Hanlon, Hroch, Hudcova, Garc{\'{i}}a, Jordan, Jordi, Kretschmar,
  Laviada, March, Mart{\'{i}}n, Mazy, Men{\'{e}}ndez, Mi, de~Miguel,
  Mu{\~{n}}oz, Nolan, Olmedo, Plesseria, Polcar, Reina, Renotte, Rochus,
  S{\'{a}}nchez, San~Mart{\'{i}}n, Smith, Soldan, Thomas, Tim{\'{o}}n, \&
  Walton}]{Mas03}
Mas-Hesse, J.~M., Gim{\'{e}}nez, A., Culhane, J.~L., {et~al.} 2003, \aap, 411,
  L261

\bibitem[{Nichols {et~al.}(2010)Nichols, Henden, Huenemoerder, Lauer, Martin,
  Morgan, \& Sundheim}]{Nichols10}
Nichols, J.~S., Henden, A.~A., Huenemoerder, D.~P., {et~al.} 2010, VizieR
  Online Data Catalog, 218, 80473

\bibitem[{Padovani \& Giommi(1995)}]{Padovani95}
Padovani, P. \& Giommi, P. 1995, \mnras, 277, 1477

\bibitem[{Perryman {et~al.}(1997)Perryman, Lindegren, Kovalevsky, Hoeg,
  Bastian, Bernacca, Cr{\'{}}~ez{\'{}} e, Donati, Grenon, van Leeuwen, van~der
  Marel, Mignard, Murray, Le~Poole, Schrijver, Turon, Arenou, Froeschl{\'{}}~e,
  \& Petersen}]{Perryman97}
Perryman, M. A.~C., Lindegren, L., Kovalevsky, J., {et~al.} 1997, \aap, 323,
  L49

\bibitem[{Piquard {et~al.}(2001)Piquard, Halbwachs, Fabricius, Geckeler,
  Soubiran, \& Wicenec}]{Piquard01}
Piquard, S., Halbwachs, J.~L., Fabricius, C., {et~al.} 2001, \aap, 373, 576

\bibitem[{Pojmanski(1997)}]{Pojmanski97}
Pojmanski, G. 1997, \actaa, 47, 467

\bibitem[{{Samus} {et~al.}(2012){Samus}, {Durlevich}, \& {et al.}}]{Samus09}
{Samus}, N.~N., {Durlevich}, O.~V., \& {et al.} 2012, GCVS database, CDS B/gcvs

\bibitem[{Scargle(1982)}]{Scargle82}
Scargle, J.~D. 1982, \apj, 263, 835

\bibitem[{Stellingwerf(1978)}]{Stellingwerf78}
Stellingwerf, R.~F. 1978, \apj, 224, 953

\bibitem[{Ubertini {et~al.}(2003)Ubertini, Lebrun, Di~Cocco, Bazzano, Bird,
  Broenstad, Goldwurm, La~Rosa, Labanti, Laurent, Mirabel, Quadrini, Ramsey,
  Reglero, Sabau, Sacco, Staubert, Vigroux, Weisskopf, \&
  Zdziarski}]{Ubertini03}
Ubertini, P., Lebrun, F., Di~Cocco, G., {et~al.} 2003, \aap, 411, L131

\bibitem[{Udalski {et~al.}(1992)Udalski, Szymanski, Kaluzny, Kubiak, \&
  Mateo}]{Udalski92}
Udalski, A., Szymanski, M., Kaluzny, J., Kubiak, M., \& Mateo, M. 1992, \actaa,
  42, 253

\bibitem[{Vedrenne {et~al.}(2003)Vedrenne, Roques, Sch{\"{o}}nfelder, Mandrou,
  Lichti, von Kienlin, Cordier, Schanne, Kn\"~odlseder, Skinner, Jean, Sanchez,
  Caraveo, Teegarden, von Ballmoos, Bouchet, Paul, Matteson, Boggs, Wunderer,
  Leleux, Weidenspointner, Durouchoux, Diehl, Strong, Cass{\'{}}~e, Clair, \&
  Andr{\'{}}~e}]{Vedrenne03}
Vedrenne, G., Roques, J.~P., Sch{\"{o}}nfelder, V., {et~al.} 2003, \aap, 411,
  L63

\bibitem[{Veron-Cetty \& Veron(1996)}]{Veron96}
Veron-Cetty, M.~P. \& Veron, P. 1996, A Catalogue of quasars and active nuclei:
  European Southern Observatory (ESO), |c1996, 7th ed.

\bibitem[{Veron-Cetty \& Veron(1998)}]{Veron98}
Veron-Cetty, M.~P. \& Veron, P. 1998, A Catalogue of quasars and active nuclei:
  8th ed., Publisher: Garching: European Southern Observatory (ESO), 1998,
  Series: ESO Scientific Report Series vol no: 18,

\bibitem[{Walker {et~al.}(2003)Walker, Matthews, Kuschnig, Johnson, Rucinski,
  Pazder, Burley, Walker, Skaret, Zee, Grocott, Carroll, Sinclair, Sturgeon, \&
  Harron}]{Walker03}
Walker, G., Matthews, J., Kuschnig, R., {et~al.} 2003, \pasp, 115, 1023

\bibitem[{Watson {et~al.}(2012)Watson, Henden, \& Price}]{Watson12}
Watson, C., Henden, A.~A., \& Price, A. 2012, VizieR Online Data Catalog, 1,
  2027

\bibitem[{Watson(2006)}]{Watson06}
Watson, C.~L. 2006, Society for Astronomical Sciences Annual Symposium, 25, 47

\bibitem[{Winkler {et~al.}(2003)Winkler, Courvoisier, Di~Cocco, Gehrels,
  Gim{\'{e}}nez, Grebenev, Hermsen, Mas-Hesse, Lebrun, Lund, Palumbo, Paul,
  Roques, Schnopper, Sch{\"{o}}nfelder, Sunyaev, Teegarden, Ubertini, Vedrenne,
  \& Dean}]{Winkler03}
Winkler, C., Courvoisier, T. J.~L., Di~Cocco, G., {et~al.} 2003, \aap, 411, L1

\bibitem[{Wo{\'z}niak {et~al.}(2004)Wo{\'z}niak, Vestrand, Akerlof, Balsano,
  Bloch, Casperson, Fletcher, Gisler, Kehoe, Kinemuchi, Lee, Marshall, McGowan,
  McKay, Rykoff, Smith, Szymanski, \& Wren}]{Wozniak04}
Wo{\'z}niak, P.~R., Vestrand, W.~T., Akerlof, C.~W., {et~al.} 2004, \aj, 127,
  2436

\bibitem[{Zacharias {et~al.}(2010)Zacharias, Finch, Girard, Hambly, Wycoff,
  Zacharias, Castillo, Corbin, DiVittorio, Dutta, Gaume, Gauss, Germain, Hall,
  Hartkopf, Hsu, Holdenried, Makarov, Martinez, Mason, Monet, Rafferty, Rhodes,
  Siemers, Smith, Tilleman, Urban, Wieder, Winter, \& Young}]{zacharias10}
Zacharias, N., Finch, C., Girard, T., {et~al.} 2010, \aj, 139, 2184

\bibitem[{Zacharias {et~al.}(2004)Zacharias, Monet, Levine, Urban, Gaume, \&
  Wycoff}]{Zacharias04}
Zacharias, N., Monet, D.~G., Levine, S.~E., {et~al.} 2004, in BAAS, Vol.~36,
  American Astronomical Society Meeting Abstracts, 1418

\bibitem[{Zwintz {et~al.}(2000)Zwintz, Weiss, Kuschnig, Gruber, Frandsen, Gray,
  \& Jenkner}]{Zwintz00}
Zwintz, K., Weiss, W.~W., Kuschnig, R., {et~al.} 2000, \aaps, 145, 481

\end{thebibliography}

\begin{appendix}
\section{Charts provided for each object in the catalogue}

Some illustrative examples of the charts provided with the catalogue for different kinds of
objects. The information about type of object, variability type and spectral type on the
top of each plot comes from Simbad and VSX. In all cases we show the DSS red image of the
field of view in the top left panel (red circles represent the OMC photometric aperture),
and the OMC V-band light curve without folding in the top right one. The time is the Barycentric INTEGRAL Julian Date expressed in Barycentric Dynamical Time (TDB). To convert this time to Barycentric Julian Date expressed in TDB, you have to add 2,451,544.5 d. Similar charts and the complete light curves in machine readable format for all the sources contained in the OMC--VAR Catalogue can be retrieved from 
\newline
\textit{http://sdc.cab.inta-csic.es/omc/}.

\begin{figure*}[th]
\centering
\vspace{0.5cm}
\includegraphics[angle=270,width=0.7\textwidth]{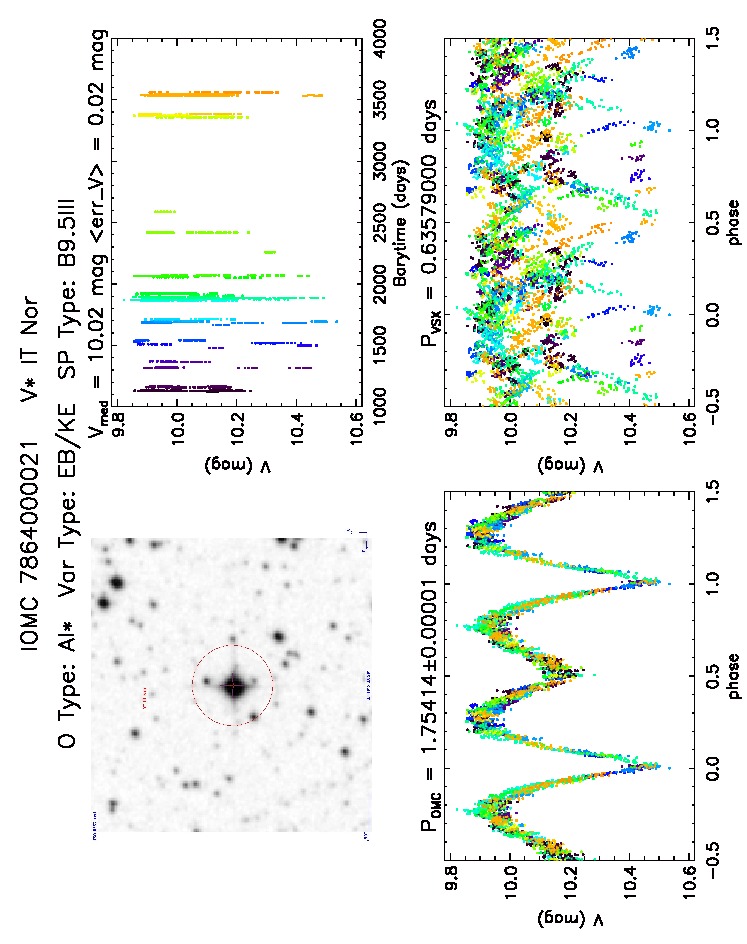}
\caption{IT Nor, an Algol type eclipsing binary. When not indicated otherwise, we show in the different charts the OMC light curve folded with the period estimated in this work (bottom left), and the OMC light curve folded with the period given by the VSX catalogue (bottom right panel). In this case the period derived from the OMC light curve improves significantly the value provided by the VSX and yields a satisfactory folding. }
\label{examples}
\end{figure*}

\begin{figure*}[!h]
\centering
\vspace{0.3cm}
\includegraphics[angle=270,width=0.7\textwidth]{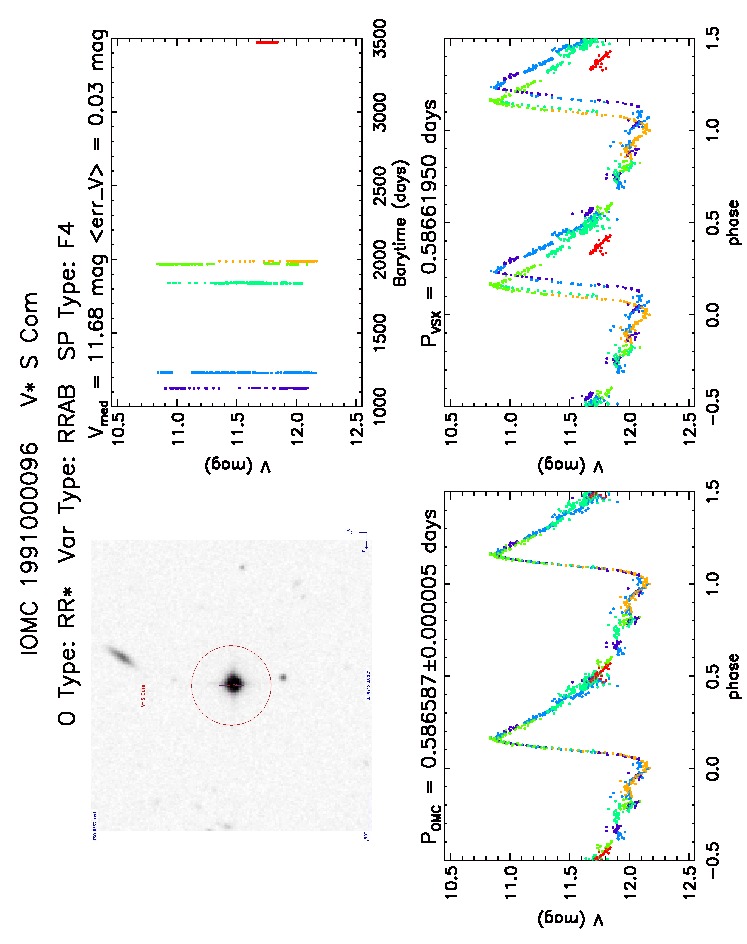}
\caption{For some objects, like the RR Lyr star S Com, a very small increase in the accuracy of the period is enough to get a perfect folding of the light curve.}

\end{figure*}

\begin{figure*}[th]
\centering
\vspace{0.5cm}
\includegraphics[angle=270,width=0.7\textwidth]{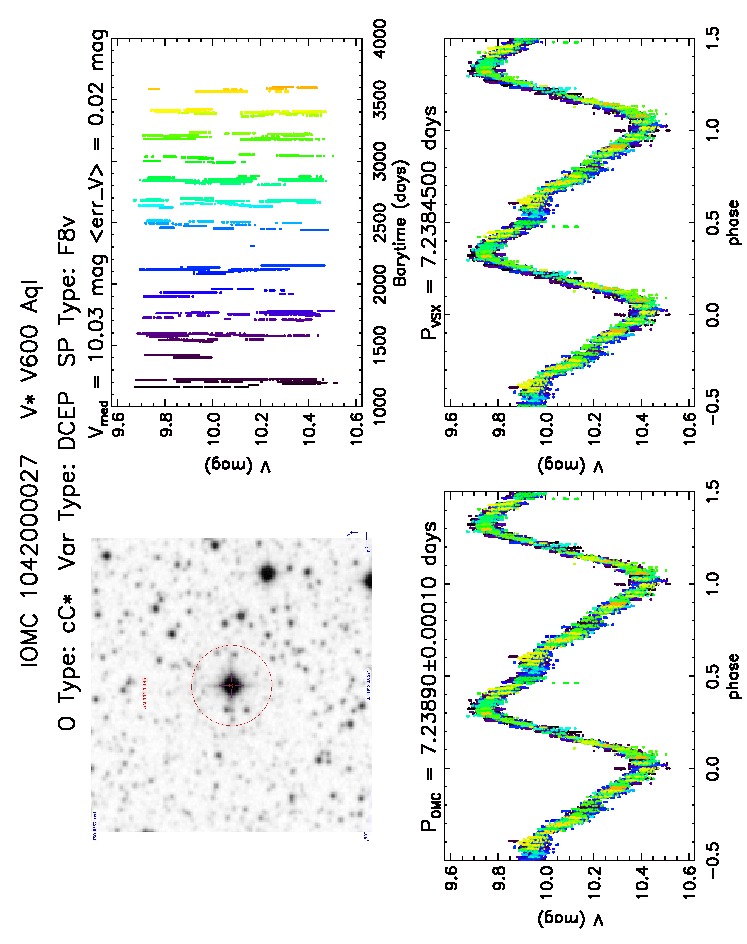}
\caption{For a large fraction of objects, the period derived from OMC data and that compiled by VSX yield similar results when folding the light curve. }
\end{figure*}

\begin{figure*}[ht]
\centering
\vspace{0.3cm}
\includegraphics[angle=270,width=0.7\textwidth]{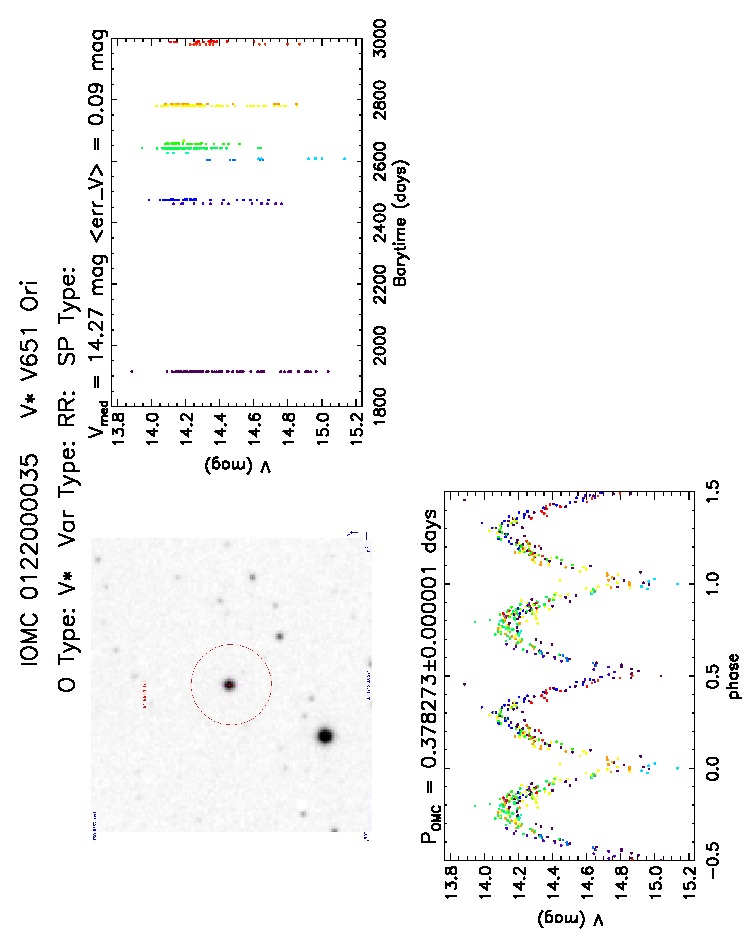}
\caption{New period determination from this work for V651 Ori. Moreover, the variability classification in VSX should be corrected from RR Lyrae to eclipsing binary, according to our light curve. }
\end{figure*}

\begin{figure*}[ht]
\centering
\vspace{0.3cm}
\includegraphics[angle=270,width=0.7\textwidth]{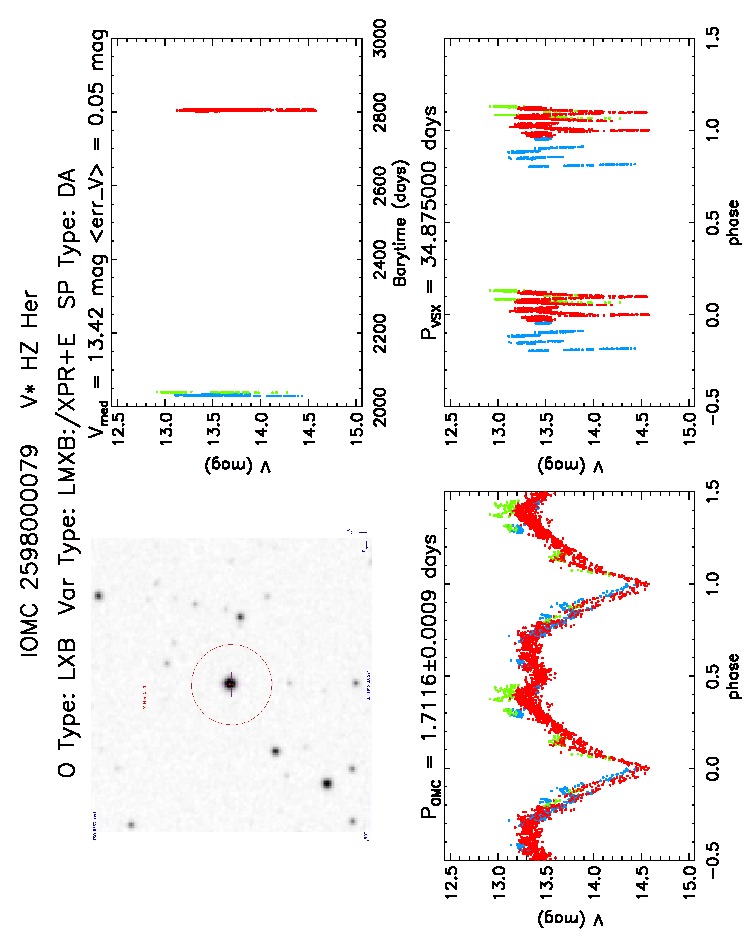}
\caption{In the case of Her~X--1, the VSX period corresponds to the 35 day cycle in the X-ray intensity, while the OMC optical light curve is dominated by the orbital eclipses.}
\label{period1}
\end{figure*}

\begin{figure*}[th]
\centering
\vspace{0.5cm}
\includegraphics[angle=270,width=0.7\textwidth]{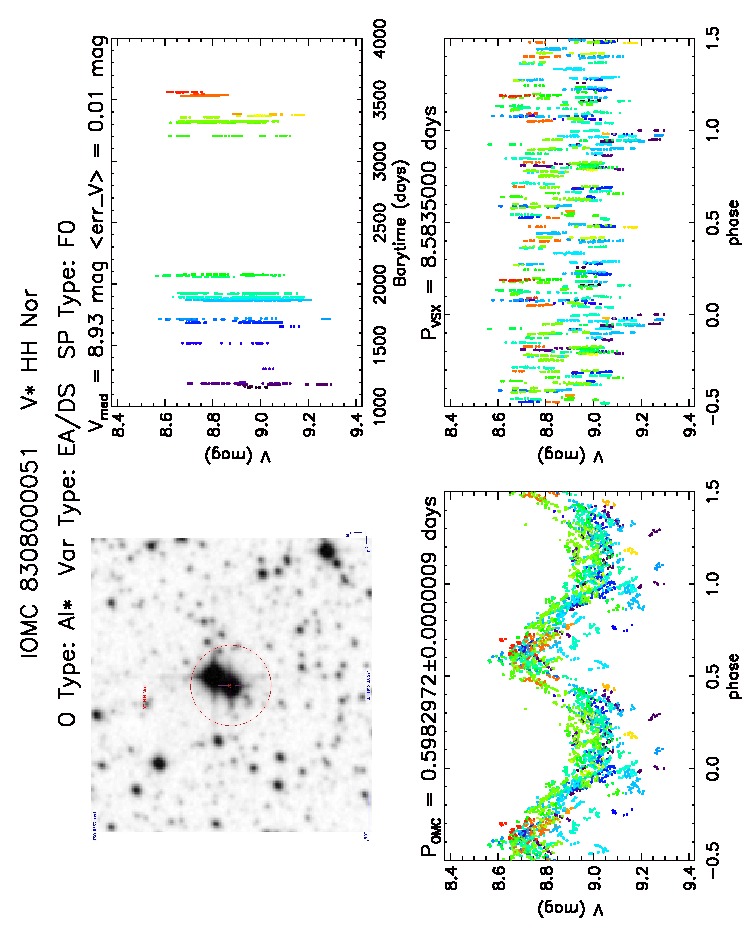}
\caption{The photometry of the intended target, HH Nor, is heavily dominated by the brighter, nearby object VSX J154329.4--515037, which is indeed a variable RR Lyr star (RRAB) with a  period of 0.59 days. In fact, HH Nor is an eclipsing binary with mean $V$ magnitude of 10.9 and a period of 8.58 days.}
\label{period2}
\end{figure*}

\begin{figure*}[th]
\centering
\vspace{0.5cm}

\includegraphics[angle=270,width=0.7\textwidth]{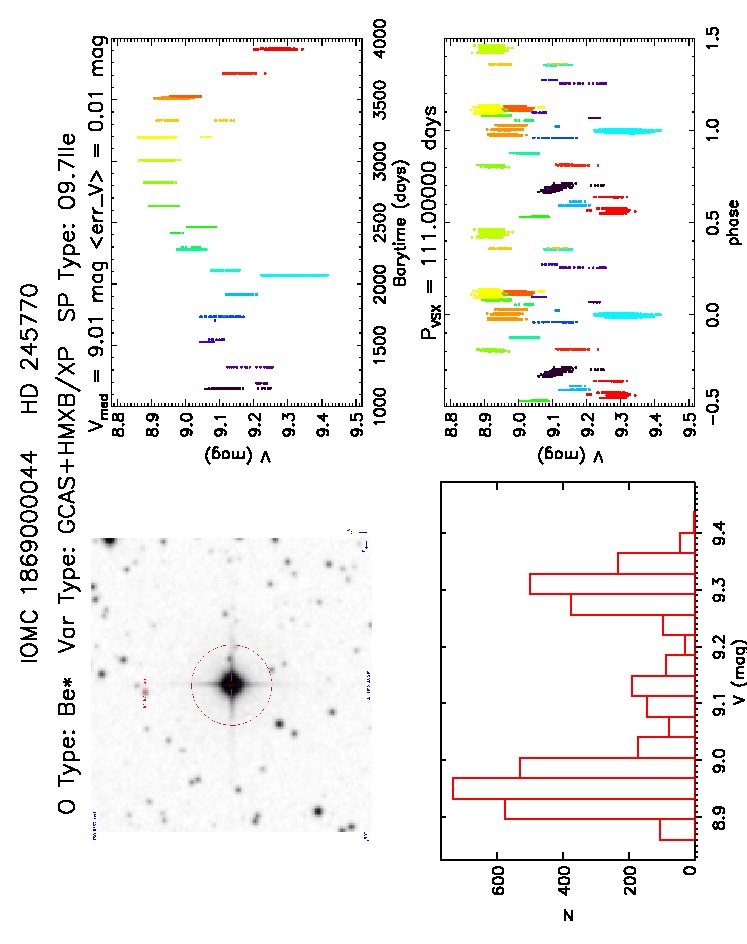}

\caption{ In the light curve of this high-mass X-Ray binary we see that a long term variation during the nearly 3000 days of monitoring dominates over the orbital period given by VSX. When there is no determination of the period from this work, the chart of the corresponding object includes a histogram of the observed magnitudes. 
The shape of these histograms helps to understand the variability pattern of the different objects.}
\end{figure*}

\begin{figure*}[ht]
\centering
\vspace{0.3cm}

\includegraphics[angle=270,width=0.7\textwidth]{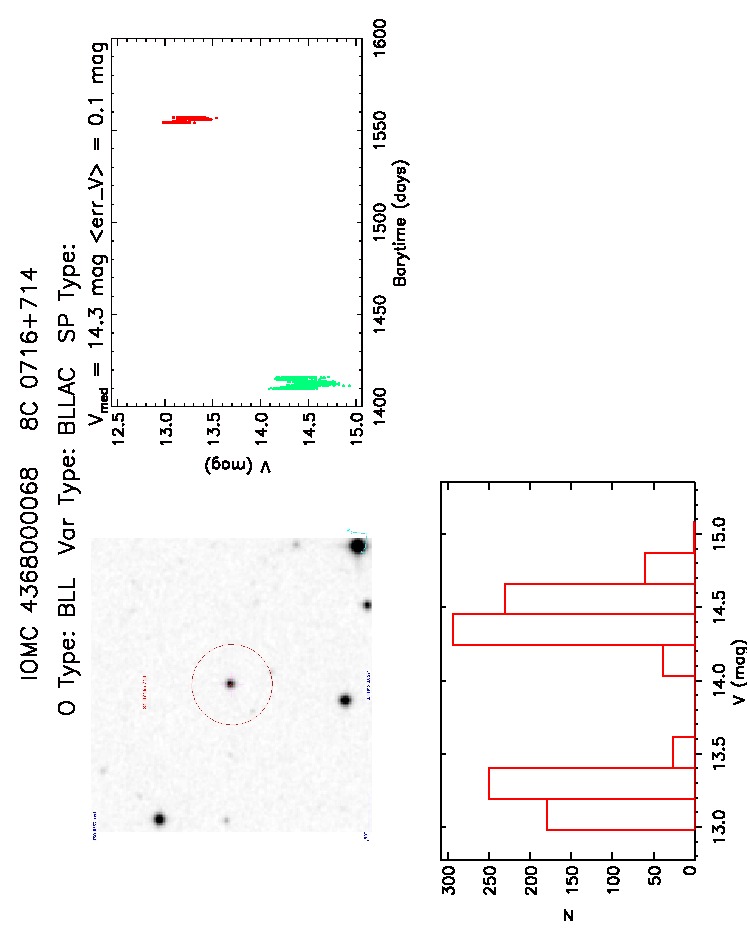}

\caption{The OMC light curve of this blazar shows two different states, separated by around 150 days. 
They can be clearly observed in the magnitude histogram.}
\end{figure*}

\end{appendix}

\end{document}